\documentclass{aa}

\usepackage{graphicx}
\usepackage{natbib}
\bibpunct{(}{)}{;}{a}{}{,}

\title{
  Dust emission model of Lyman-break galaxies
}

\author{
  T.~T.~Takeuchi\inst{1}\thanks{
  Postdoctoral Fellow of the Japan Society for the Promotion of Science for 
  Research Abroad.
}
  \and 
  T.~T.~Ishii\inst{2}\thanks{
  Postdoctoral Fellow of the Japan Society for the Promotion of Science.
}
}

\institute{Laboratoire d'Astrophysique de Marseille, 
  Traverse du Siphon, BP8-13376 Marseille Cedex 12, France\\
  \email{tsutomu.takeuchi@oamp.fr}
\and
  Kwasan Observatory, Kyoto University, Yamashina-ku, Kyoto,
  607--8471, Japan\\
  \email{ishii@kwasan.kyoto-u.ac.jp}
}

\offprints{T. T. Takeuchi\\
\email{tsutomu.takeuchi@oamp.fr}.
}

\date{Received/Accepted}

\def\sbs{SBS~0335$-$052}
\def\izw{I~Zw~18}
\def\cb{MS~1512$-$cB58}

\titlerunning{Dust Emission Model of LBGs}
\authorrunning{T.\ T.\ Takeuchi \& T.\ T.\ Ishii}

\begin{document}

\abstract{
Lyman-break galaxies (LBGs) contain a non-negligible amount of dust.
\citet{takeuchi03a} (T03) constructed a model of the infrared spectral 
energy distribution (SED) for very young galaxies by taking into 
account the dust size distribution in the early stage of galaxy evolution,
which can be different from that of present-day evolved galaxies.
We applied T03 model to LBGs and constructed their expected SED.
In order to examine the grain size distribution of dust, we calculated
the SEDs based on two distinct type of the distribution models: 
a single-sized distribution and a power-law distribution with a slope of 
$dN/da \propto a^{-3.5}$.
We found that the single-sized and power-law dust size distributions yield a
very similar detectability of LBGs at the submillimetre (submm).
We also found that galaxies with a power-law dust distribution have 
much less flux at mid-infrared (MIR) than the other.
By making use of this fact we can explore the dust grain size distribution 
in high-redshift galaxies through (observer-frame) FIR observations.
Then, we applied the model to a gravitationally lensed LBG \cb\ (cB58),
a unique probe of the dust emission from LBGs.
Observations by SCUBA suggest that the dust is hot in this galaxy.
Our model framework well reproduced the hot dust temperature under a natural
assumption for this galaxy.
We also examined the detectability of LBGs at submm wavelengths
in an eight-hour deep survey by ALMA.
The LBG population with an age $\ga 10^8\,{\rm yr}$ and
a SFR $\ga 10\,M_\odot {\rm yr}^{-1}$ can be detected in such a survey.
By integrating over their redshifted SEDs with the observed luminosity 
functions, we obtained the contribution of LBGs to the cosmic infrared 
background radiation (CIRB).
Although they have non-negligible amount of dust, their contribution
was found to be small, especially in the FIR $\sim 200\,\mu$m.
Thus, we need a strongly obscured population of galaxies which contains
a large amount of star formation, at some epoch in the history of 
the universe.
}

\maketitle

\section{Introduction}

Star formation and metal enrichment in an early stage of 
galaxies are of great importance to understand the history of the universe.
With the aid of a variety of new observational techniques and 
large facilities, studies on such young galaxies are being pushed to higher
and higher redshifts.
The surveys that have most efficiently secured large population
of galaxies at a redshift $z\sim 3$ are those that exploit the
Lyman break to identify high-$z$ candidates: Lyman-break galaxies 
(LBGs) \citep{steidel99,steidel03}.

Even in LBGs, there is clear evidence that they contain 
non-negligible amount of dust \citep[e.g.,][]{
sawicki98,adelberger00,calzetti01,shapley01}. 
Dust grains absorb stellar light and re-emit it in the far infrared (FIR),
hence it is crucial to evaluate the intensity and spectrum of FIR 
emission from galaxies for understanding their star formation properties
\citep[e.g.,][and references therein]{buat02,hirashita03}.
Despite its importance, there still remains a large uncertainty in our
understanding of the properties of dust emission from LBGs
\citep[for a review, see][]{calzetti01}.

\citet{ouchi99} have estimated the submm flux from LBGs and, for the 
first time, found empirically that the dust temperature should be high 
in order to explain their non-detection by SCUBA.
\citet{chapman00} also pointed out that the SCUBA submm 
flux of LBGs were much weaker than that expected from their UV spectral 
slope index.
They discussed that it is hard to reconcile unless dust temperature $T$ 
is higher than that of local starbursts ($T \ga 70\,$K).
For the analysis of SCUBA flux of a highly lensed LBG, MS~1512$-$cB58 
(hereafter cB58), \citet{sawicki01} (S01) performed a thorough survey of the 
parameter space of dust temperature and emissivity index,
and concluded that the dust temperature and/or emissivity index in cB58
is substantially higher than those seen in local galaxies.

Another important issue to be addressed is the contribution of 
dust emission from LBGs to the cosmic infrared background (CIRB)
\citep{gispert00,hauser01,takeuchi01a,totani02}.
The star formation rate (SFR) density of LBGs is now believed to increase 
toward redshifts $z \ga 3$, if we `correct' dust extinction \citep{steidel99}.
Then, the re-emission from dust in LBGs is an interesting target to examine.
Some resent observations suggest that their contribution to the total CIRB
is not very large, at most $\sim 20\;$\% \citep{webb03}.
If so, there must be other contributors to the CIRB, which should bear an
intense star formation comparable or even larger than that in LBGs.
Hence, theoretical prediction of the dust emission plays an important role
in the interpretation of the cosmic SFR and the CIRB.

In order to address these issues and to make a consistent picture of 
the dust emission from LBGs, we must consider their star formation history, 
chemical evolution and dust mass evolution, radiation process with 
nonequilibrium temperature fluctuation, and dust composition properly.
Recently, \citet{hirashita02} (H02) have developed a model 
for the evolution of dust content in very young galaxies (age $t \la 10^9\,
\mbox{yr}$).
Type II supernovae (SNe II) are the dominant source for
the production of dust grains in young star-forming galaxies
\citep{dwek80}.
H02 have modeled the evolution of FIR luminosity and dust temperature 
in such a young starburst on the basis of SNe II grain formation model of 
\citet{todini01} (TF01). 

\citet{takeuchi03a} (T03) subsequently constructed a model of infrared
(IR) spectral energy distribution (SED) of galaxies starting from the H02 
model.
It has been believed that the size of dust grains formed in SNe II cannot be 
as large as $0.1 \mbox{--} 1\,\mu$m (TF01), and host galaxies are too young 
for grains to grow in the interstellar space.
T03, for the first time, properly considered the dust size distribution 
peculiar to the very early stage of galaxy evolution, and construct 
a model of the IR SED of very young galaxies.
T03 model successfully reproduced the peculiar SED of a local 
metal-deficient galaxy \sbs\ ($Z = 1/41\,Z_\odot$), which we regarded as 
an analogue of genuine young galaxies.
However, in clear contrast, recently \citet{nozawa03} has presented a new 
model of dust grain formation in SNe II, which produce larger grains 
($\sim 0.1 \mbox{--} 1\;\mu$m) than the previously believed conjecture.
Their dust has a broken power-law grain size distribution.
Based on the known physical condition in LBGs, we discuss the possible 
geometrical configuration of dust, and show that we can test the dust
size distribution at high redshift by an observation of LBGs 
in the IR and submm.
Such observations will be soon possible by, e.g.,{\sl SPICA}, {\sl Herschel}, 
SMA and ALMA.

The layout of this paper is as follows:
In Section~\ref{sec:model} we briefly describe our model framework.
We present our basic results for the SEDs of LBGs in 
Section~\ref{sec:results}.
Related discussions are given in Section~\ref{sec:discussion}.
First we focus on a gravitationally magnified LBG \cb, a unique probe of 
the internal physics of LBGs.
Then we consider some cosmologically important problems.
Section~\ref{sec:conclusion} is devoted to our summary and conclusions.

\section{SED Model for Extremely Young Galaxies}\label{sec:model}

\subsection{SED construction}

Since H02 treats the evolution of dust content in a galaxy younger than
$10^8$~yr, the only contributor to the total dust mass ($M_{\rm dust}$) in 
a young ($<10^8$~yr) galaxy is the supply from SNe II.
The rate of SNe II is given by
\begin{eqnarray}
  \gamma (t)=\int_{8M_\odot}^{\infty} \psi (t-t_m)\, \phi (m)\, dm\, ,
\end{eqnarray}
where $\psi (t)$ is the SFR at age $t$ (we define $t=0$ at the beginning of 
the star formation), $\phi (m)$ is the initial mass function (IMF), 
$t_m$ is the lifetime of a star whose mass is $m$, 
and it is assumed that only stars with $m>8~M_\odot$ produce SNe II. 
H02 assume a constant SFR, $\psi =\psi_0$, and a Salpeter IMF.
Then the rate of increase of $M_{\rm dust}$ is written as
$\dot{M}_{\rm dust}=m_{\rm dust}\gamma$ ($m_{\rm dust}$ is the typical 
dust mass produced by one SN II).
H02 adopted $m_{\rm dust}=0.4~M_\odot$. 

In general we must take into account not only the production but also 
destruction of dust grains.
However, the timescale of dust destruction is much longer than that we 
consider in this work, hence the effect of the destruction appears to be 
negligible.
Detailed evaluation can be found in H02 [their eq.~(2)].

Dust grains are roughly divided into silicate and carbonaceous ones.
TF01 have presented that the sizes of silicate and 
carbonaceous grains formed in SN II ejecta to be about 10~\AA\ and 300~\AA, 
respectively.
Classical studies claimed that the dust grains originating from SNe II 
cannot be as large as those formed in the atmosphere of evolved AGB stars or 
grown in the diffuse interstellar space (\citealt{kozasa87}; 
TF01).
In this case, the discrete and small grain sizes make the appearance of 
the IR SED of young galaxies drastically different from that of aged normal 
galaxies.
On the other hand, recently \citet{nozawa03} have proposed 
a drastically different picture of dust size distribution from SNe II.
They showed that, under some condition, dust grains can grow large 
even within the expansion timescale of SNe ejecta.
Consequently, their size distribution of grains have a broken power-law shape.
For comparison, we also consider this case by using the simple description 
of Galactic dust \citep{mathis77} as
\begin{eqnarray}\label{eq:mdust}
  \frac{dN}{da} \propto a^{-3.5} 
\end{eqnarray}
for both grain species, i.e., silicate or carbon dust.
Here we note that we do not include polycyclic aromatic hydrocarbons (PAHs)
in our dust grain model, because they are rarely found in metal-poor systems
and/or in intense radiation environments \citep[e.g.,][]{madden00,galliano03}.

It is well accepted that very small grains are stochastically heated,
that is they cannot establish thermal equilibrium with the ambient radiation 
field \citep{draine85,draine01}:
the heat capacity of very small grains is too small to maintain its
temperature until the next photon impinges.
As a result, the temperature of a grain varies violently in time, and
we can define a distribution of the instantaneous temperature of a grain, 
which will be introduced in eq.~(\ref{eq:lum_IR}).
For the heat capacity of a grain, $C(T)$, we apply the multidimensional 
Debye model according to \citet{draine01}.

Then we calculated the temperature distribution of dust 
as a function of size $a$ by Monte Carlo simulations, following 
\citet{draine85}.
The UV radiation field strength is determined by the history of OB star 
luminosity of H02 and the size of the star forming region $r_{\rm SF}$. 
For the geometry, we assume that the OB stars are concentrated 
in the centre of the system, and dust is assumed to be distributed as a shell.
We regard the radius of the sphere, $r$ as the size of the spherical 
star forming region, $r_{\rm SF}$.
The flux is expressed by $J(t)=L_{\rm OB}(t)/(4\pi r_{\rm SF}^2)$, hence
the UV energy density, $u$, is obtained by $u=J(t)/c$.
The spectral UV energy density $u_\lambda$ is calculated from the assumed 
spectrum of the UV radiation field.

The rate at which a grain absorbs a photon with energy $E \sim E+dE$
is expressed as\footnote{We note that there are two typographic 
errors in Equations~(13) and (14) of T03.
The equations we show here are the correct expressions.}
\begin{eqnarray}\label{eq:prob}
  \frac{d^2 p}{dEdt} = Q_{\rm abs} (a, \lambda)\pi a^2 u_\lambda 
    \frac{\lambda^3}{h^2c} \,,
\end{eqnarray}
where $Q_{\rm abs}$ is the absorption efficiency of a dust grain.
We used the values proposed by \citet{draine84} for $Q_{\rm abs}$ of 
silicate and carbon grains.

The heating is represented as follows:
\begin{eqnarray}\label{eq:heating}
  \frac{hc}{\lambda} = \frac{4\pi a^3}{3}\int_{T_0}^{T} C(T')dT' \, ,
\end{eqnarray}
where $T$ is the peak temperature achieved by a grain hit by a 
photon with energy $hc/\lambda$, and $T_0$ is the grain temperature 
just before absorption.   
On the other hand, the grain cools through radiation as 
\begin{eqnarray}\label{eq:cooling}
  \frac{d T}{dt} = -\frac{3\pi}{aC(T)} 
    \int_0^\infty Q_{\rm abs}(a,\lambda)\, B_\lambda(T) d\lambda \,,
\end{eqnarray}
where $B_\lambda(T)$ is the blackbody function represented as a function
of wavelength.

{}From Equations~(\ref{eq:prob}), (\ref{eq:heating}), and (\ref{eq:cooling}),
we obtain the sample path of a grain temperature as a function of time.

The total mass of each grain component is given by TF01.
The mass ratio we adopt here is $M_{\rm sil}:M_{\rm C} = 0.56:0.44$
(H02).
With this value and material density of each species ($\rho_{\rm
sil}=3.50\,\mbox{g}\,\mbox{cm}^{-3}$ and $\rho_{\rm
C}=2.26\,\mbox{g}\,\mbox{cm}^{-3}$: \citealt{draine84}), 
we obtain the normalization of the dust size distribution $dN_i/da_i'$,
\begin{eqnarray}\label{eq:dust_number_norm}
  M_{\rm dust} f_i =
    \int \frac{4\pi {a_i'}^3 \rho_i}{3} \frac{dN_i}{da_i'}da_i' \,,
\end{eqnarray}
where subscript $i$ denotes the species of dust, sil or C, and $f_i$ 
is the mass fraction of dust of species $i$.
The total number of grains, $N_i$, is expressed as 
\begin{eqnarray}\label{eq:dust_number}
  N_i = \int \frac{dN_i}{da_i'}da_i' \,.
\end{eqnarray}
For the TF01 dust distribution, $a_i$ is fixed for each dust component to
a specific value $a_i$, i.e., ${dN_i}/{da_i'}=N_i\delta(a_i'-a_i)$.
In this case, with Equations~(\ref{eq:dust_number}) and 
(\ref{eq:dust_number_norm}), the normalization reduces to 
\begin{eqnarray}\label{eq:dust_number_single}
  N_i = \frac{3M_{\rm dust}f_i}{4\pi a_i^3\rho_i} \,.
\end{eqnarray}
We drop the prime in the following expressions.

Total IR emission spectrum from a galaxy is then calculated by 
superposing each of the continuum from silicate and carbonaceous grains,
\begin{eqnarray}\label{eq:lum_IR}
  L_{{\rm IR},\lambda}(t) =
    \sum_i \pi \int\int 4 \pi a_i^2 
    Q_{\rm abs}^i(\lambda)B_\lambda(T)
    \frac{dN_i}{da_i}\frac{dP_i(u,a_i)}{dT}\,dT da_i \,,
\end{eqnarray}
where $dP_i(u,a_i)/dT$ is the temperature probability density function,
i.e., the fraction of time for which a grain with species $i$ in a UV 
radiation field with energy density $u$ stays in a temperature range 
$[T, T+dT]$.
The starburst age is incorporated with $N_i$ (i.e., $dN_i/da_i$) and 
$dP_i(u,a_i)/dT$ through $M_{\rm dust}(t)$ and $u_\lambda (t)$.
Here we put a constraint that the amount of absorbed light is equal to
the FIR luminosity.
Lastly, if the dust opacity is very large, then self-absorption occurs, and 
even MIR radiation from dust is absorbed by the dust itself.
We treat the extinction as given by a shell model. 
We crudely approximate the dust opacity and extinction to the first order as
\begin{eqnarray}\label{eq:tau}
  \tau_{\rm dust}(\lambda) &\simeq&
    \sum_i \pi a_i^2 Q_{{\rm abs},i} (\lambda)\,
    \frac{3 N_i}{4\pi \left[(r_{\rm SF}+\Delta r_{\rm SF})^3 
    - r_{\rm SF}^3\right]} \Delta r_{\rm SF} \nonumber \\
  &\simeq& \sum_i \pi a_i^2 Q_{{\rm abs},i} (\lambda)\,
    \frac{N_i}{4\pi r_{\rm SF}^2}\,, \\
  I(\lambda) &=& 
    I_0(\lambda) e^{-\tau_{\rm dust}(\lambda)} \,.
\end{eqnarray}
The absorbed light is re-emitted at longer wavelengths, mainly in the 
submm, and consequently, the SED is deformed by the self-absorption.
Final SED is obtained via this absorption--re-emission process.
For further technical details, see T03.

\subsection{Input parameters for LBGs}\label{subsec:input}

Here we consider the input physical parameters in the model calculation for 
LBGs.
We adopted the dust grain sizes $a_{\rm C}=200\,\mbox{\AA}$ and 
$a_{\rm sil}=6\,\mbox{\AA}$ for our canonical model.
They are the same as those for \sbs\ and \izw\ used in T03, because LBGs
are thought to be young, and hence we expect similar dust properties to them.

The SFR of LBGs spreads over the range of 
$\mbox{SFR} \simeq 1 \mbox{--} 300\,M_\odot \mbox{yr}^{-1}$ with a median 
value of $\mbox{SFR} \simeq 20\,M_\odot \mbox{yr}^{-1}$ 
\citep[e.g., ][]{erb03}.
We assumed a constant SFR up to the age shorter than $10^9$~yr, 
which may be a good approximation \citep[e.g.,][]{baker04}.
Within the young age of $t \la 10^9\,$yr, neither SNe~I nor RGB/AGB stars 
contribute to the dust production.
Other dust growth mechanism in the interstellar medium cannot work
effectively, either \citep[see, e.g.,][]{whittet92}.
Thus, the basic framework of T03 model is valid for LBGs.

The most important information to calculate the IR SED is the effective
size of the star forming region, but it is the most uncertain quantity
at the same time.
It is still beyond the possibility of the present facilities to 
measure the size directly from observations.
The mean half-light radius of LBGs is estimated to be $\sim 1.6\,
\mbox{kpc}$ from {\sl HST} observations \citep{erb03}, 
hence it may be a general value for LBG populations.
The similarity between the rest UV and optical morphologies suggests that 
the dust in LBGs may not preferentially obscure particular regions in these 
galaxies \citep{dickinson00,calzetti01}.
Then we can safely use the galaxy radius $r_{\rm G}$ as the radius of 
a star-forming region, $r_{\rm SF}$, i.e., $r_{\rm SF} = 2\,\mbox{kpc}$.
We will revisit this issue with more physical considerations in
Section~\ref{sec:discussion}.

\section{Results}\label{sec:results}

\begin{figure*}
\centering\includegraphics[angle=90,width=17cm]{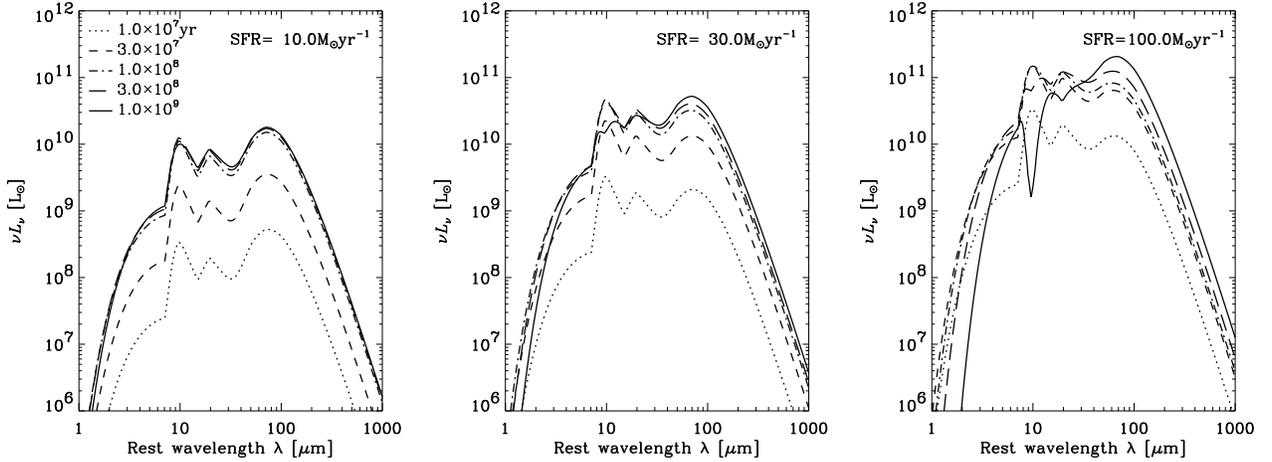}
\caption{The evolution of the SED for the single-sized dust size distribution.
{}From left to right, the star formation rate (SFR) is 10, 30, and 
$100\,M_\odot\mbox{yr}^{-1}$.
The dotted, dashed, dot-dashed, long-dashed, and solid lines represent
that the age of the major star formation in a galaxy is $1.0\times 10^7\,$yr,
$3.0\times 10^7\, $yr,$1.0\times 10^8\,$yr, $3.0\times 10^8\,$yr, 
$3.0\times 10^7\,$yr, and $1.0\times 10^9\,$yr, respectively.
}\label{fig:sed_lbg_sdust}
\end{figure*}

\begin{figure*}
\centering\includegraphics[angle=90,width=17cm]{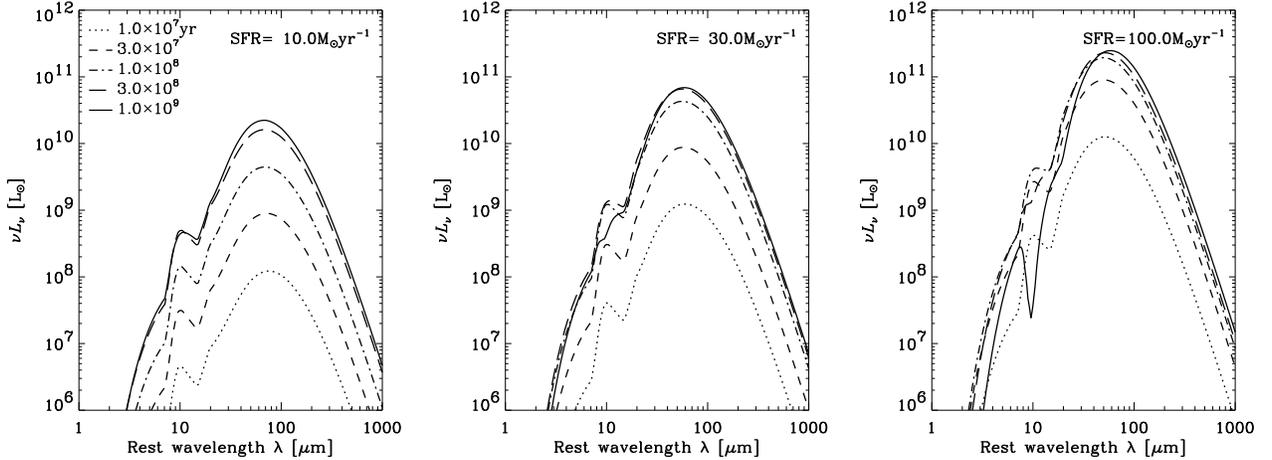}
\caption{The evolution of the SED for the power-law dust size distribution.
Comparing with Figure~\ref{fig:sed_lbg_sdust}, we see a clear difference 
in the MIR.
}\label{fig:sed_lbg_mdust}
\end{figure*}

Based on the above-mentioned settings, we calculated the IR SEDs of LBGs
with $\mbox{SFR} = 10$, 30, and $100\,M_\odot\mbox{yr}^{-1}$ as 
representative cases.
We traced the evolution of the SEDs from $t=10^7\,\mbox{yr}$ to 
$10^9\,\mbox{yr}$.

Figure~\ref{fig:sed_lbg_sdust} shows the evolution of the SED for the 
single-sized dust size distribution.
The dust opacity is found to be not very large, and consequently, 
the silicate features appeared in emission.
Only in the case of $\mbox{SFR}=100\,M_\odot\mbox{yr}^{-1}$, these features
becomes a significant absorption after $3 \times 10^8\,\mbox{yr}$.
Most of the LBGs have a lower SFR than $\mbox{SFR}=100\,M_\odot
\mbox{yr}^{-1}$, hence basically we can expect the silicate emission band 
feature in their MIR spectra.
Along the same line, since the self-absorption is not significant for LBG 
spectra, we also expect a strong MIR continuum emission from
LBGs if the single-sized dust distribution takes place.

For comparison, we next show the evolution of the SED for the 
power-law dust size distribution in Figure~\ref{fig:sed_lbg_mdust}.
In this case, an obvious difference is that the MIR radiation is an order of
magnitude weaker than that for the single-sized dust results.
This is because the power-law size distribution has much smaller amount of 
small grains ($a \la 100\,$\AA) than the single-sized distribution.
The extinction properties are similar to that of single-sized dust case,
despite the drastic difference of the SEDs.
Thus, if the geometrical configuration of dust is similar among LBGs, 
the MIR observation can be a strong tool to explore the size distribution
of dust grains.

\section{Discussions}\label{sec:discussion}

\begin{figure*}
\centering\includegraphics[angle=90,width=17cm]{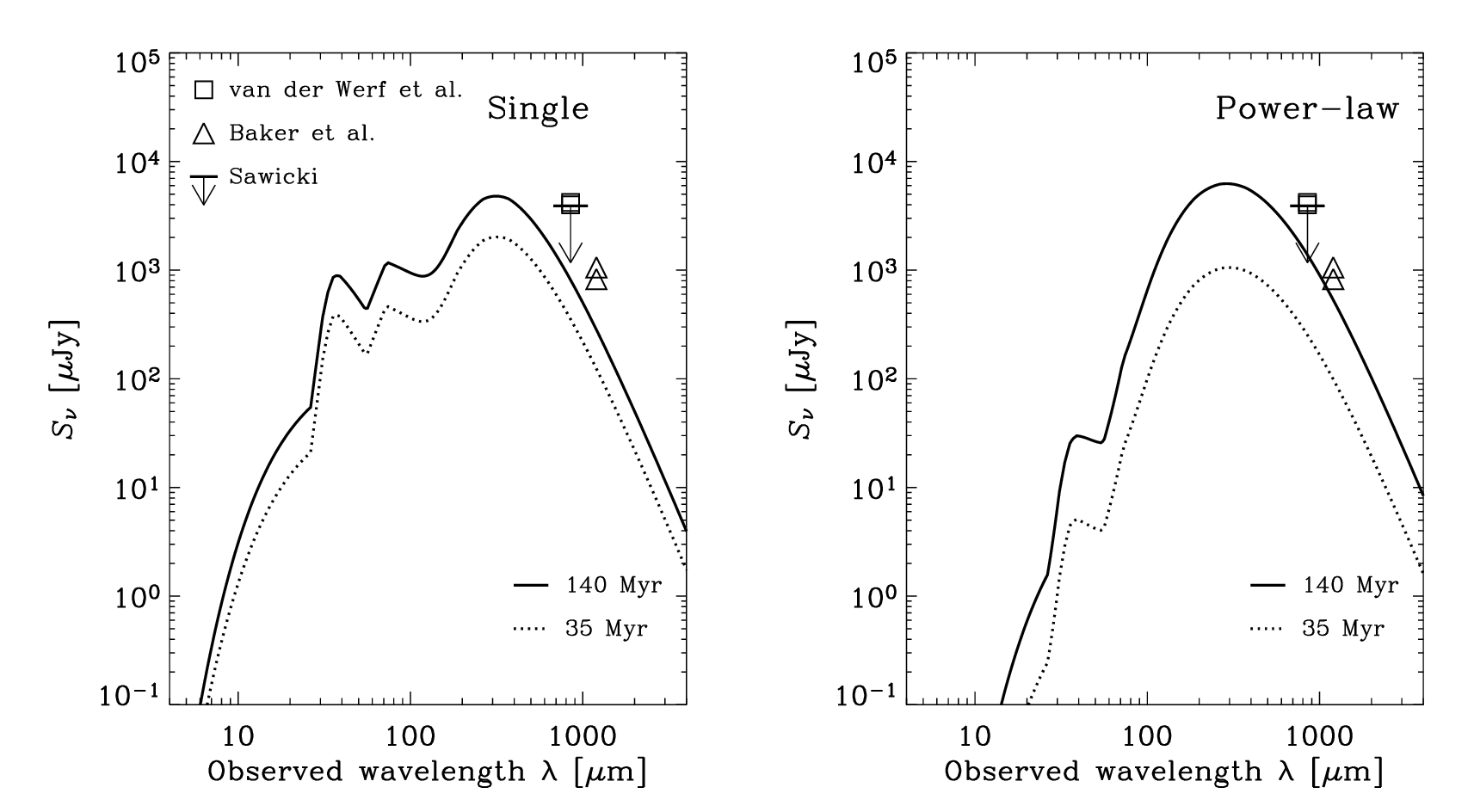}
\caption{Model spectral energy distribution (SED) of a lensed Lyman-break
galaxy \cb\ (cB58).
Left panel shows the observed SED calculated based on the single-sized dust 
size distribution with silicate grain radius 
$a_{\rm sil}=6\,\mbox{\AA}$ and carbon grain radius $a_{\rm C} = 
200\,\mbox{\AA}$.
The observed SED is magnified by a factor 30.
Right panel shows the SED with a power-law grain size distribution.
Symbols represent the measured flux densities reported by
\cite{sawicki01} (upper limits), \cite{baker01} (open triangles) and 
\cite{vanderwerf01} (open squares).
In both panels, Dotted lines represent the SED with the age of 
$t_{\rm SF}=35\,$Myr, and dashed lines of $t_{\rm SF}=140\,$Myr, 
respectively.
}\label{fig:sed_cb}
\end{figure*}

\subsection{\cb}\label{subsec:cb}

It is still difficult to compare a model prediction for young galaxies to 
observational data of LBGs.
Among the LBGs, a typical $L_*$ LBG cB58 is a fortunate exception:
it has been observed in various wavelengths 
\citep[e.g.,][]{ellingson96,bechtold97,nakanishi97,teplitz00,baker01}, 
because of a strong magnification by 
gravitational lensing \citep[factor $22\mbox{--}40$:][]{seitz98}. 
We adopt a commonly used value of 30 for magnification.
The half-light radius of cB58 is estimated to be
$r_{\rm G} \simeq 1.4h^{-1}\,$kpc from a detailed image reconstruction
\cite{seitz98}\footnote{Throughout this paper, we adopt a cosmology with
$(h,\Omega_0,\lambda_0)=(0.7,0.3,0.7)$, where $h\equiv H_0/100.$}.

{}From spectroscopic observations of nebular lines, the SFR of cB58 is
estimated to be $\simeq 10 \mbox{--}20h^{-2}\, M_\odot \mbox{yr}^{-1}$
\citep{pettini00}. 
The age of major star formation in cB58 was estimated to be $t_{\rm SF} 
\simeq 35\,\mbox{Myr}$ \citep{matteucci02}. 
An older age estimate of $t_{\rm SF} = 140\,$Myr is given by 
\citep{baker04}, and we adopt both of these values for our calculations.
This is broadly consistent with other observational suggestions 
\citep[e.g.,][]{pettini00,ellingson96}, and significantly younger than 
the median value of those of general LBGs 
\citep[$\sim 350\,$Myr:][]{shapley01}.

The optical extinction of cB58 is estimated to be $E(B-V) \simeq 0.27$
\citep{pettini00,ellingson96,teplitz00}.
Emission from dust in cB58 is measured at two wavelengths:
\citet{sawicki01} observed this galaxy at $850\,\mu$m by SCUBA on 
the JCMT, and no signal was detected above $3\sigma$ level of 3.9~mJy.
Another observation by SCUBA \citep{vanderwerf01} detected a $850\,\mu$m 
flux to be $4.2\pm 0.9$~mJy.
\citet{baker01} reported a detection at 1.2~mm by MAMBO (Max-Planck 
Millimeter Bolometer) array, and the flux was $1.06 \pm 0.35$~mJy 
($4.4\sigma$).

The submm flux from the foreground cD galaxy (6$''$ away from cB58)
may contaminate the observed data \citep{sawicki01}. 
After careful examination of the new submm mapping data, it has been 
concluded that the contamination is
$\sim 5$~\% at $850\,\mu$m and $\sim 22$~\%  at 1.2~mm, respectively
\citep{baker04}, though there still remain some other
uncertainties (e.g., \citealt{baker04} discusses a possible contribution of 
emission lines or background sources. 
We may need an interferometric observation to solve this problem).
In addition, the uncertainties in the lensing correction and in the UV based
SFR estimation may also be as large as that of the contamination correction.
We should keep these issues in mind in the following discussions.

\subsubsection{IR SED}\label{subsubsec:irsed}

We use a commonly accepted value of $24\,M_\odot \mbox{yr}^{-1}$ as a SFR
\citep{baker04} and assume the SFR to be constant in time.
For the age of the major star formation in cB58, as we mentioned above, 
we adopt $35\,$Myr and $140\,$Myr.
Because of its young age, dust is predominantly produced by SNe II and 
the framework of T03 model is again valid for cB58.
The half-light radius of the reconstructed image of cB58 
\citep[$\simeq 2\,\mbox{kpc}$][]{seitz98} is very similar to that of general 
LBGs \citep[$\sim 1.6\, \mbox{kpc}$][]{erb03}.

We show the model SED of cB58 in Figure~\ref{fig:sed_cb}.
Left panel presents the SED with a single-sized dust, and 
Right panel shows the SED with a power-law distribution
(Equation \ref{eq:mdust}), respectively.
Symbols represent the measured flux densities reported by
\citet{sawicki01} (upper limits), \citet{baker01} (open triangles) and 
\citet{vanderwerf01} (open squares). 
There are two open symbols at each wavelength: the upper open symbols are 
the measured values, and the lower ones represent the flux densities 
corrected for the possible contamination by the point source at the lens 
cluster center \citep{baker04}.
It is still not clear whether the point source is the cD galaxy or not
\citep{baker04}.

The apparent peak of the SED is located at a wavelength shorter than 
$100\,\mu$m at the restframe of cB58. 
Actually, the dust temperature is higher than that of the local 
($z\sim 0$) sample of normal galaxies observed by SCUBA, 
but comparable to the most intense dusty starbursts \citep{dunne01}. 
It is mainly because of an intense UV radiation field of cB58 and low dust
opacity. 
The mean dust temperature of the power-law dust model is lower than that of
the single-sized dust model, but the superposition makes the submm
part of its final SED similar to that of a hot dust emission. 
We find that both models predict slightly lower flux densities than observed.
Considering the uncertainties mentioned above, we conclude that the model is
roughly consistent with observations within a few factors.
Some authors claim that the submm flux densities of cB58 is surprisingly
low \citep[e.g.,][]{baker04}, but in our framework, they are naturally
explained by our model framework (the model predicts even smaller fluxes).
The power-law model predicts a slightly larger flux at submm wavelengths, 
but it is difficult to distinguish the two models only by submm observations.

The prominent feature in the MIR is the silicate band of $9.7\,\mu$m.
For the single-sized dust model, the NIR--MIR part of the SED is dominated 
by the radiation from small silicate grains, and FIR--submm part by larger
carbonaceous grains.
In contrast, relatively lower abundance of small grains in the power-law 
model makes the MIR radiation considerably weaker.

\subsubsection{Extinction and size of the star-forming regions}
\label{subsec:extinction_cb}

\begin{figure}
\resizebox{\hsize}{!}{
\includegraphics{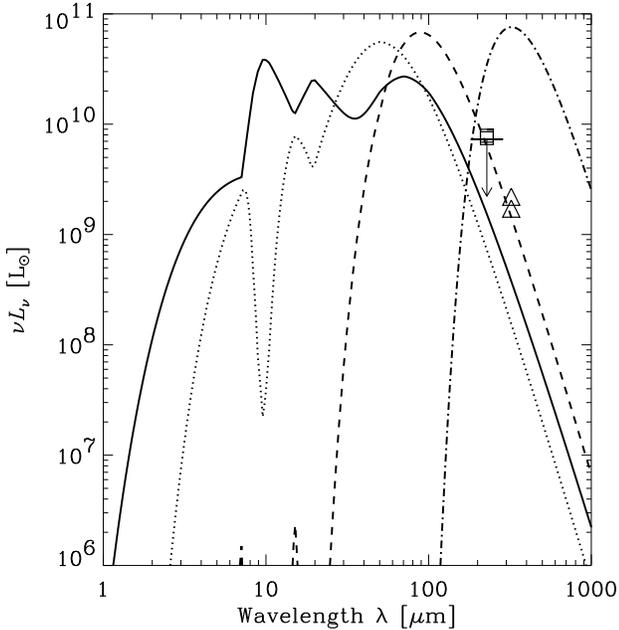}
}
\caption{Effect of the size of the star-forming regions on the model SED
of cB58.
Solid line is the canonical model (same as the solid line in the left panel 
of Fig.~\ref{fig:sed_cb}).
Dotted, dashed, and dot-dashed lines indicate the SED calculated
with a radius of the star-forming region, $r_{\rm SF}=300\,\mbox{pc}$, 
120\,pc, and 30\,pc, respectively.
Note that the SED is presented in the unit of $\nu L_\nu\,[L_\odot]$ here.
}\label{fig:sed_sfr_size}
\end{figure}

Besides IR emission, dust extinction is a fundamental quantity
to specify the dust properties. 
We relate the extinction in magnitude ($A_\lambda$) with 
$\tau_{\rm dust} (\lambda)$ (Equation~\ref{eq:tau}) by 
\begin{equation}
  A_\lambda = 1.086 \tau_{\rm dust}(\lambda)\,.
\end{equation}
When we adopt the $r_{\rm SF}=2\,\mbox{kpc}$, the extinction at $K$-band
is $A_K =0.08\,\mbox{mag}$.
Again we adopt the extinction curve of \citet{cardelli89}, we expect
$A_V =0.8\,\mbox{mag}$ from our model.
The measured color excess is $E(B-V)=0.27$, hence by adopting $R_V =3.1$ as a 
representative value in a diffuse ISM \citep{krugel03}, 
we have $A_V = R_V E(B-V) \simeq 0.84\,\mbox{mag}$, 
which is in perfect agreement with our model prediction.
\citet{vijh03} performed a sophisticated analysis of the attenuation in LBGs 
by using a radiative transfer model.
They showed the attenuation is in the range of $5.7\mbox{--}18.5$ at $1600\,
\mbox{\AA}$,
and is well represented by Calzetti law \citep{calzetti94}.
Our extinction estimate is also consistent with their result.
They also showed that the shell geometry is appropriate for the dust in LBGs,
which is in accordance with our assumption.
Thus, we see that the extinction in the visible stellar system of cB58
is consistent with its dust amount and spatial scale of 
$r_{\rm SF}\simeq 2\,\mbox{kpc}$.

Then, can a heavily hidden star formation exist in this galaxy?
Actually, \sbs\ hosts a completely extinguished starburst
\citep{hunt01,plante02}, and its SED is well understood by T03 model.
{}To consider this problem, we have calculated the SEDs with various
smaller star-forming region sizes.
This means that the bulk of the dust in cB58 could be confined in 
a small part of its volume.
They are shown in Figure~\ref{fig:sed_sfr_size}.
Dotted, dashed, and dot-dashed lines are those with $r_{\rm SF}=300\,$pc,
120\,pc, and 30\,pc, respectively.
Smaller sizes make the opacity in the region significantly larger, and
the conversion from UV to FIR becomes much more efficient.
Considering the observational upper limits, the allowed smallest size is 
$r_{\rm SF}=120\,$pc, with a possible range of $50\mbox{--}200\,\mbox{pc}$  
if we take into account the uncertainties in the estimates of the SFR 
and lensing magnification.

It has been pointed out that the galactic wind can transport the dust
into larger radius (T03).
Actually, precise spectroscopy by \cite{pettini03} clearly shows the 
outflowing motion of gas with a velocity of $v_{\rm wind} \simeq 
255\,\mbox{km\,s}^{-1}$ in cB58.
They consider the wind and extinction, and suggest that
the superwind shell size is comparable to the galaxy size.
Hence the wind crossing timescale is $t_{\rm cross} \sim 10^6\,\mbox{yr}$.
If dust closely couples with the gas motion, it is difficult to
keep dust grains confined in a region much smaller than the galaxy size
through its age of $t = 35\mbox{--}140\,\mbox{Myr}$.
Strong radiative pressure may also expand the dusty region, as suggested by 
\citep{inoue02}.
Here we should comment the effect of the patchiness of dust, 
which may be realistic for LBGs.
In the FIR, since the opacity of dust is small ($\tau < 1$),
the FIR emission escapes from a galaxy without significant absorption.
Then the flux is controlled simply by energy conservation, and 
the patchiness does not affect the result significantly
\citep[see, e.g.,][]{witt00,gordon00}.

Together with its high SFR and SN rate, which are considered to 
provide a large amount of radiative and kinetic energy to gas-dust system, 
we conclude that a compact dusty star formation may not be plausible for cB58.
Hence, the dust configuration of cB58 may be similar to the present model.
This suggests that we can safely discriminate the effect of dust size 
distribution on the SED from that of the configuration and geometry of dust
in the galaxy.
Again, this conclusion can hardly be affected by the clumpiness of dust, 
just the same reason mentioned above.

\subsubsection{Follow-up observations of cB58}
\label{subsec:followup_cb}

Our single-sized dust model provides an SED with bright continuum in the 
rest NIR--MIR regime, hence we expect that cB58 will be detected by
MIR--FIR observations ($\lambda \sim 20\mbox{--}100\,\mu$m).
At present, {\sl Spitzer Space Telescope}\footnote{URL: 
{\tt http://sirtf.caltech.edu/.}} MIPS and
{\sl ASTRO-F}\footnote{URL: 
{\tt http://www.ir.isas.ac.jp/ASTRO-F/index-e.html.}} 
IRC will be suitable instruments for such observations.
For example, both instruments have the same level of sensitivity
around $\sim 20~\mu$m, and the flux of 0.06--0.08 mJy will be detected.
For FIR, both can detect 0.8--1 mJy at $\lambda =60\mbox{--}70\;\mu$m.
Therefore, the MIR continuum can be detected by the future
MIR--FIR observations if the single-sized dust takes place

On the other hand, if the power-law distribution takes place, the flux
densities will be an order of magnitude smaller than those of our canonical 
model, because the number of the stochastically heated grains is
reduced.
As we mentioned, the uncertainty caused by dust geometry is not very large
for a LBG because the configuration becomes more or less shell-like in these
galaxies just as a scale-up of \sbs. 
Therefore, the MIR observation will be a strong test for the size distribution.

We should note, however, that even by {\sl Spitzer}, confusion limit can 
be severe in FIR \citep[e.g.,][]{ishii02,dole03,takeuchi04}.
Larger space IR facilities such as 
{\sl Herschel Space Observatory}\footnote{URL: {\tt 
http://www.rssd.esa.int/herschel/.}} or 
{\sl SPICA}\footnote{URL: {\tt http://www.ir.isas.ac.jp/SPICA/index.html.}}, 
and IR interferometry missions are longed to improve the confusion limit.
At longer wavelengths, of course ALMA will be useful.

\subsection{Observability of the dust emission from LBGs}
\label{subsec:observability}

Considering the non-negligible extinction in LBGs, their energy budget
radiated in the IR should be significant \citep{adelberger00,calzetti01}.
Stimulated by the consideration, the relation between LBGs and dusty 
starbursts, another representative star-forming galaxy population at 
$z=2\mbox{--}5$, has gathered the researchers' attention.
In fact, however, most of the LBGs have not been detected by SCUBA 
observations \citep{chapman00,webb03}, except for a few extreme cases
\citep[e.g.,][]{chapman02}.
\citet{calzetti00} pointed out the possibility that the dust is hot in
LBGs based on the observed SED of a low-metallicity galaxy Tol~1924$-$416.
Hot dust in LBGs was also suggested by S01 from his observation of a lensed
LBG cB58.
Since their consideration is empirical, and we need some explanation for 
the hot dust, if it exists.
As seen in the above, our model naturally predict a high dust temperature.
Here, we discuss how LBGs look like in the IR/submm wavelengths.
It is also useful to estimate the observational feasibility of LBGs
by forthcoming submm facilities like ALMA.

\begin{figure*}
\centering\includegraphics[angle=90,width=17cm]{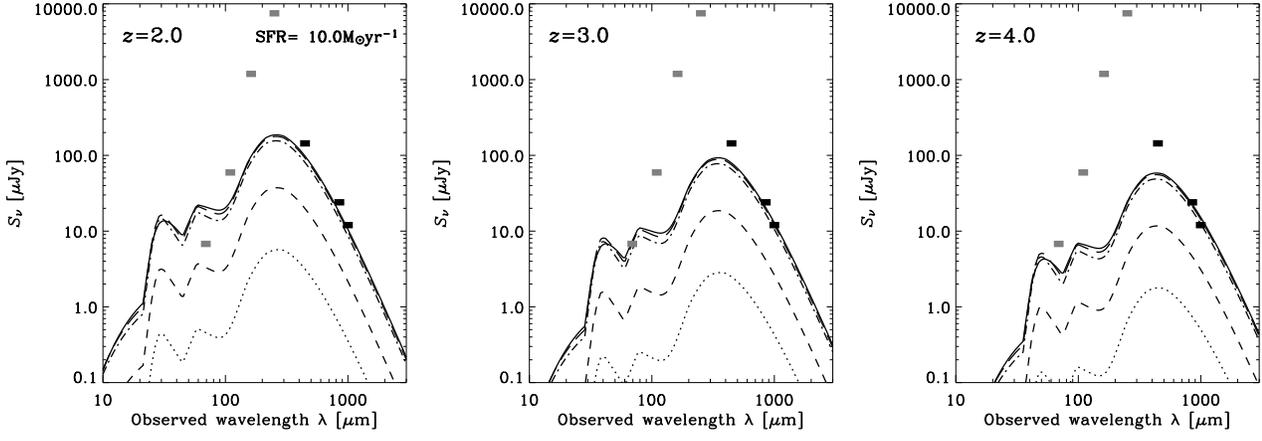}
\caption{Prediction for the observed IR/submm SEDs of LBGs at $z=2$, 3, and 4.
The star formation rate (SFR) is $10\,M_\odot \,{\rm yr}^{-1}$, and the dust
grains are single-sized.
The dotted, dashed, dot-dashed, long-dashed, and solid lines 
represent, the same as Fig.~\ref{fig:sed_lbg_sdust}, 
that the age of the major star formation in a galaxy is $1.0\times 10^7\,$yr,
$3.0\times 10^7\, $yr,$1.0\times 10^8\,$yr, $3.0\times 10^8\,$yr, 
$3.0\times 10^7\,$yr, and $1.0\times 10^9\,$yr, respectively.
Confusion limits of {\sl Herschel} and detection limits of ALMA 8-hour survey
are also shown by gray and black thick horizontal lines, respectively.
}\label{fig:sed_lbg_sdust_sfr010_obs}
\end{figure*}
\begin{figure*}
\centering\includegraphics[angle=90,width=17cm]{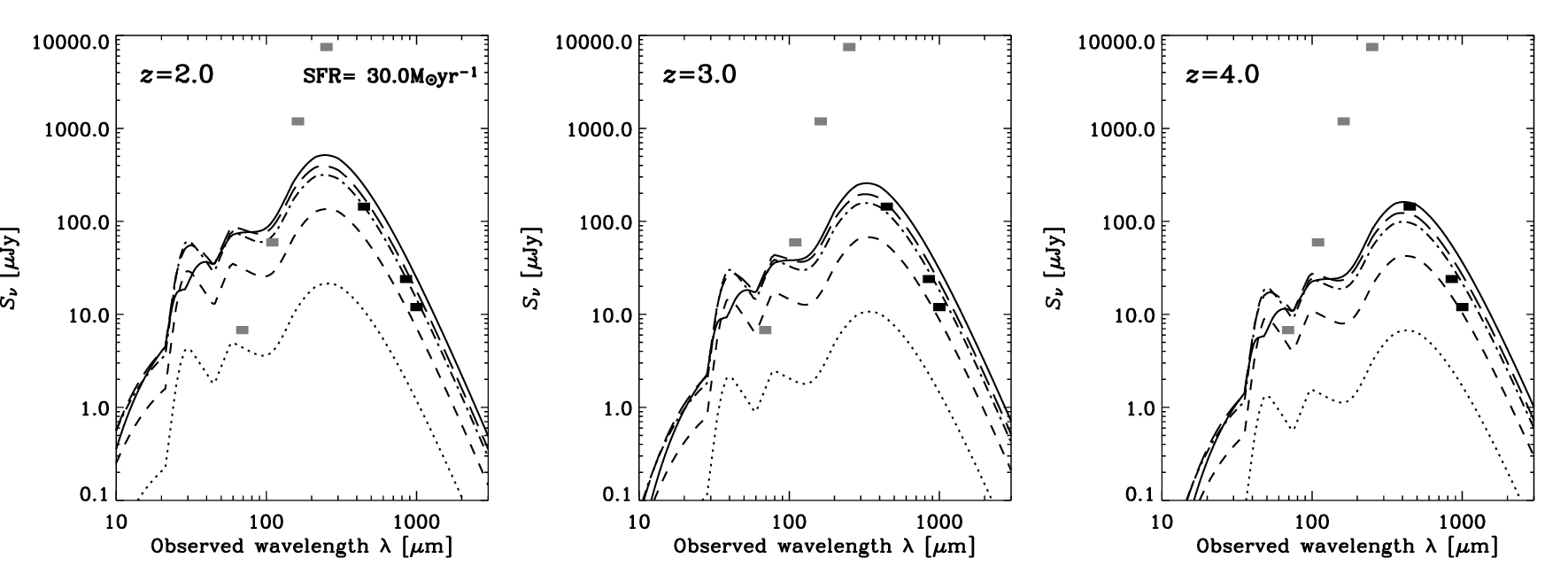}
\caption{The same as Figure~\ref{fig:sed_lbg_sdust_sfr010_obs}, but the SFR
is $30\,M_\odot \,{\rm yr}^{-1}$.
}\label{fig:sed_lbg_sdust_sfr030_obs}
\end{figure*}
\begin{figure*}
\centering\includegraphics[angle=90,width=17cm]{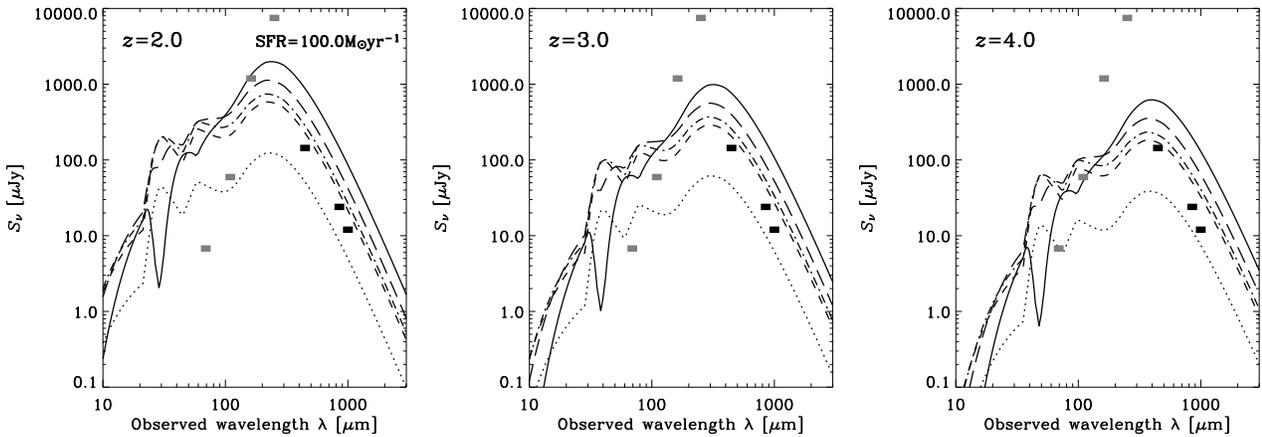}
\caption{The same as Figure~\ref{fig:sed_lbg_sdust_sfr010_obs}, but the SFR
is $100\,M_\odot \,{\rm yr}^{-1}$.
The silicate feature finally appears as absorption.
}\label{fig:sed_lbg_sdust_sfr100_obs}
\end{figure*}
\begin{figure*}
\centering\includegraphics[angle=90,width=17cm]{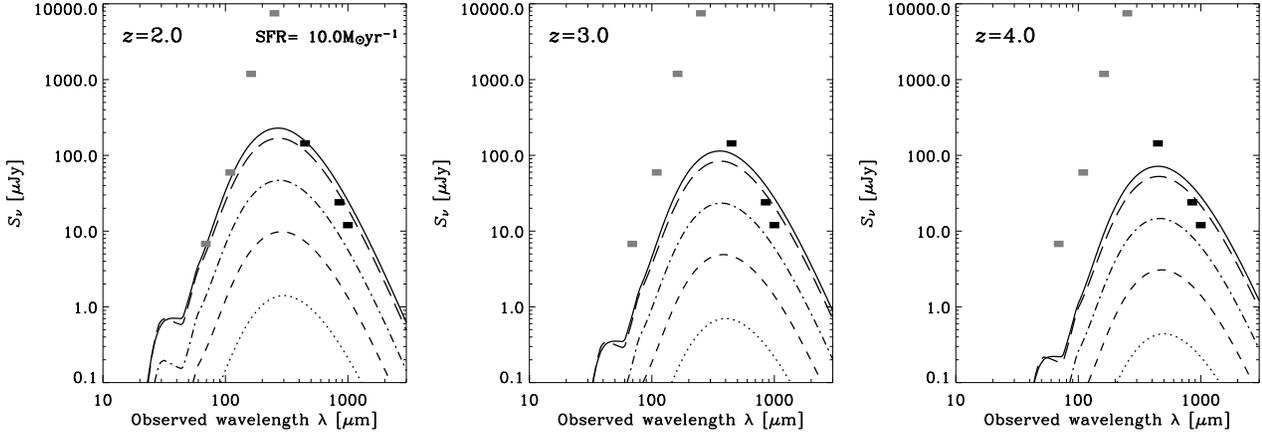}
\caption{Prediction for the observed IR/submm SEDs of LBGs at $z=2$, 3, and 4.
The star formation rate (SFR) is $10\,M_\odot \,{\rm yr}^{-1}$, and the dust
grains are power-law.
}\label{fig:sed_lbg_mdust_sfr010_obs}
\end{figure*}
\begin{figure*}
\centering\includegraphics[angle=90,width=17cm]{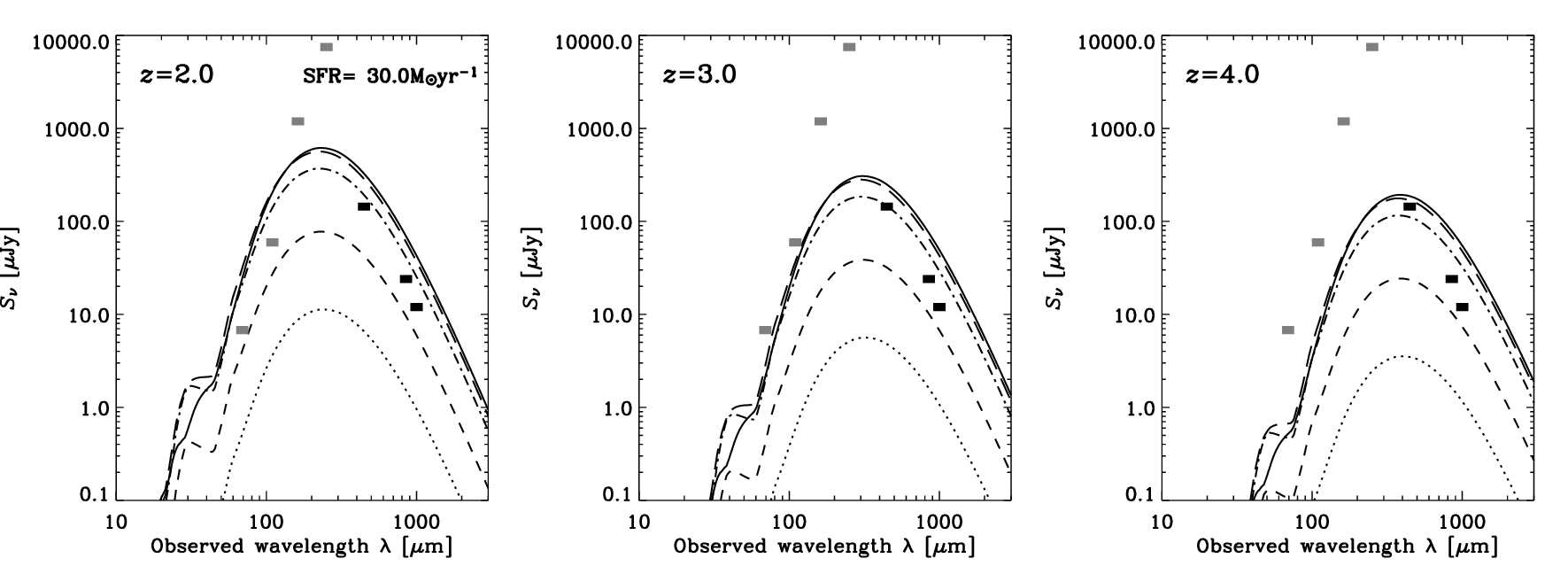}
\caption{The same as Figure~\ref{fig:sed_lbg_mdust_sfr010_obs}, but the SFR
is $30\,M_\odot \,{\rm yr}^{-1}$.
}\label{fig:sed_lbg_mdust_sfr030_obs}
\end{figure*}
\begin{figure*}
\centering\includegraphics[angle=90,width=17cm]{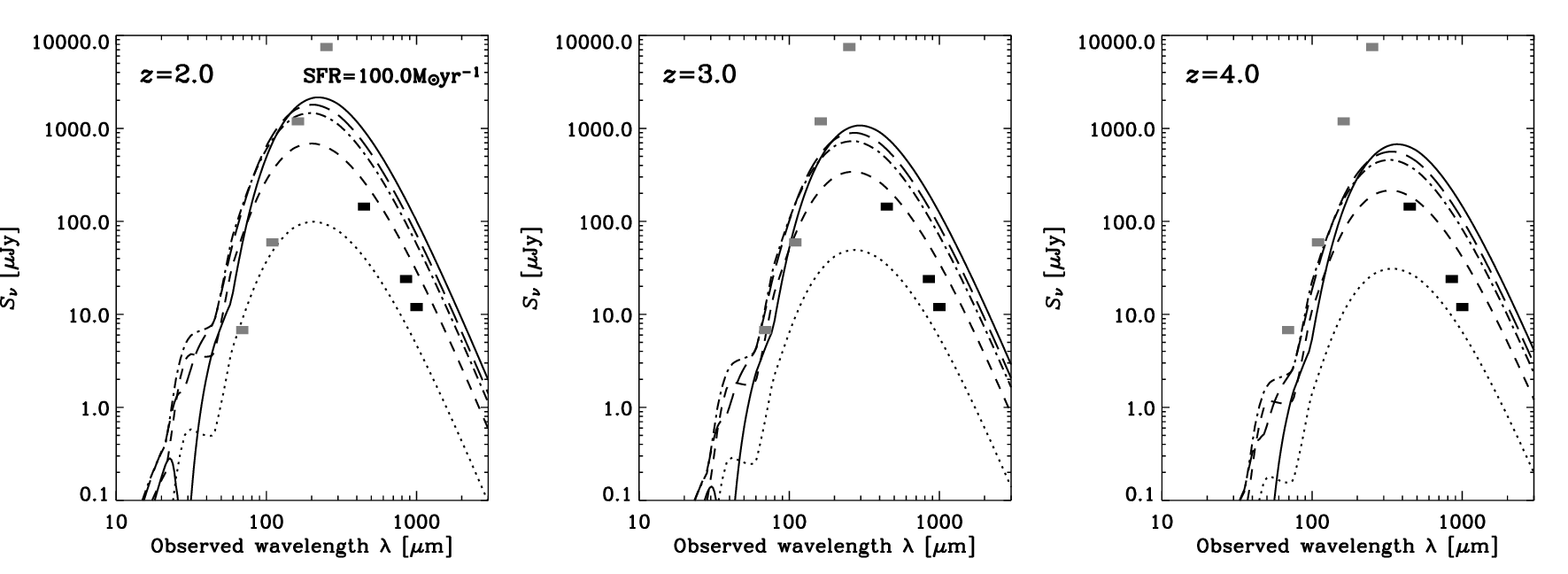}
\caption{The same as Figure~\ref{fig:sed_lbg_mdust_sfr010_obs}, but the SFR
is $100\,M_\odot \,{\rm yr}^{-1}$.
}\label{fig:sed_lbg_mdust_sfr100_obs}
\end{figure*}

The observed flux density of a source at observed frequency, $\nu_{\rm obs}$
is obtained by 
\begin{eqnarray}\label{eq:flux}
  S_{\nu_{\rm obs}} =
    \frac{(1+z)L_{(1+z)\nu_{\rm obs}}}{4\pi d_{\rm L}(z)^2} 
    = \frac{(1+z)L_{\nu_{\rm em}}}{4\pi d_{\rm L}(z)^2} \,,
\end{eqnarray}
where $d_{\rm L}(z)$ is the luminosity distance corresponding to a 
redshift $z$, and $\nu_{\rm obs}$ and $\nu_{\rm em}$ are observed and emitted
frequency, respectively.
We show the observed IR/submm SEDs of LBGs at $z=2$, 3, and 4 in 
Figures~\ref{fig:sed_lbg_sdust_sfr010_obs}--\ref{fig:sed_lbg_mdust_sfr100_obs}.
Figures~\ref{fig:sed_lbg_sdust_sfr010_obs}--\ref{fig:sed_lbg_sdust_sfr100_obs}
represent the SEDs for the single-sized dust size distribution, 
and 
Figures~\ref{fig:sed_lbg_mdust_sfr010_obs}--\ref{fig:sed_lbg_mdust_sfr100_obs}
for the power-law distribution (Equation~\ref{eq:mdust}).
The thick black short horizontal lines indicate the 3-$\sigma$ detection limits
for 8-hour observation by ALMA.
Here we assumed 64 antennas and three wavelength bands, 450, 850, and 
$1080\,\mu$m.
We also show the 3-$\sigma$ source confusion limit of {\sl Herschel} 
at 75, 160, 250, and 350$\,\mu$m bands by thick gray horizontal lines.
These limits are based on `the photometric criterion' of \citep{lagache03}.
\citet{ishii02} also report similar estimates for the confusion limits.\footnote{
Generally, in a sky region suitable for cosmological surveys, the cirrus 
confusion is less severe than the galaxy confusion
\citep[see, e.g.,][]{dole03}.}

{}From 
Figures~\ref{fig:sed_lbg_sdust_sfr010_obs}--\ref{fig:sed_lbg_mdust_sfr100_obs},
we recognize that the detectability of these galaxies 
at submm bands are not significantly dependent on the redshift.
This is because of the `negative $K$-correction', well-known in the field of 
submm astronomy.
On the contrary, the galaxy age and SFR are more important for the detection 
of LBGs.
If the age $\ga 10^8\,{\rm yr}$ and $\mbox{SFR} \ga 10\,M_\odot\,
{\rm yr}$, a LBG can be detected at a wide range of redshifts in the submm.

In general, the longer the wavelength is, the easier the detection becomes
for millimetre (mm) observations.
In the submm and mm wavelengths, the detectability is not quite different
between LBGs with single-sized dust and those with power-law dust.
It is a natural consequence that the radiation in these wavelengths is
dominated by relatively large dust grains that can establish a thermal
equilibrium with the radiation field.

The apparent peak of the dust emission lies in a relatively short wavelength
of about $200\mbox{--}400\,\mu$m at $z=2\mbox{--}4$.
Thus, the accumulated background radiation spectrum from LBGs may have 
its peak in the FIR.
The expected dust emission from LBGs is, however, too faint to be detected
even by {\sl Hershel} deep survey, except some rare exceptionally bright 
sources, e.g., lensed galaxies like \cb, or in the case of extremely high SFR
(in order to reach the confusion limits, unrealistic integration time might 
be required).
Hence, still more powerful instruments at these wavelengths would be required.

In the shorter wavelengths (MIR--FIR), the size distribution affects the
possibility of detection drastically, just as we have seen in the case of
cB58 (\ref{subsec:cb}).
Surveys in the MIR by {\sl Spitzer} and {\sl ASTRO-F} will be desirable to 
detect these objects.
These observations are important to compare the overall SEDs of LBGs directly 
with those of ``local analogues'', such as \sbs\ and \izw.
Existence of such local galaxies with hot dust is reported by
recent studies \citep[see e.g.,][]{calzetti00,takeuchi03b}.
Similarity and difference between them provides us a unique clue to 
the physics of galaxy formation.

\subsection{Contribution of LBGs to the CIRB}

\begin{figure}
\resizebox{\hsize}{!}{
\includegraphics{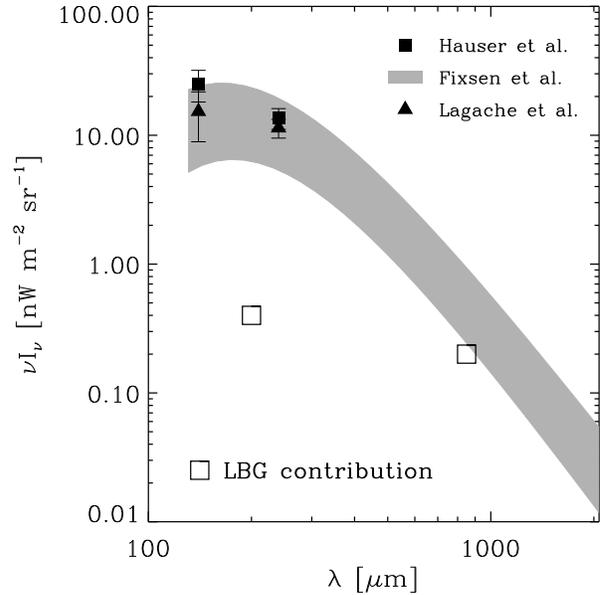}
}
\caption{
The contribution of LBGs to the cosmic infrared background radiation (CIRB)
(open symbols).
Filled symbols and hatched area are measured CIRB spectrum by {\sl COBE} 
\citep{hauser98,fixsen98,lagache99}.
}\label{fig:cirb}
\end{figure}

The implication of hot dust also offers a potentially important
cosmological insight into high-$z$ galaxies: if dust in high-$z$ object is 
as cool as usually assumed, then we cannot expect a high comoving SFR
density because in a realistic cosmology the cosmic IR background (CIRB) 
spectrum strongly constrains the high-$z$ IR emission \citep{takeuchi01a}.
However, very hot dust in such high-$z$ object reconciles the CIRB 
constraint with high SFR, and allows vigorous star formation hidden by dust 
in the early Universe \citep{totani02}.

Actually, non-detection of LBGs in the submm wavelengths 
\citep{chapman00,webb03} suggests the hot dust temperature.
Considering the non-negligible extinction in these galaxies, their energy 
budget radiated in the IR is significant \citep{adelberger00,calzetti01}.
However, their contribution to the submm radiation cannot be quite large
because of the high dust temperature. 
In fact, the contribution to the extragalactic submm background radiation has
been suggested to be at most $\sim 20$\% \citep{webb03}.

Using our model SED and the luminosity function of LBGs \citep{shapley01}, 
we calculate the integrated IR light from LBGs to the CIRB. 
In order to have a rough estimate, we correct the extinction to the observed
luminosity function, and converted it to the distribution of the SFR by 
converting the UV luminosity to the SFR according to our model.
Here we crudely assumed that all the LBGs have the same star forming region size 
$r_{\rm SF} = 2\,\mbox{kpc}$.
We also assumed that the number of LBGs rapidly decreased at $z \ga 4$.\footnote{
Actually, this assumption affect the results little.}
By using the SFR distribution, we obtained the multiband FIR luminosity function of 
the LBGs.
We finally integrated it over the FIR luminosity to get the CIRB intensity.

Our model predicts $0.4\,\mbox{nW\,m}^{-2}\mbox{sr}^{-1}$ 
at $200\,\mu$, and $0.2\,\mbox{nW\,m}^{-2}\mbox{sr}^{-1}$ at $850\,\mu$m, 
respectively.
The difference of the dust size distribution does not affect the CIRB spectrum
almost at all.
This result is shown in Figure~\ref{fig:cirb}.
The $850\,\mu$m intensity is quite consistent with the LBG contribution at 
$850\,\mu$m estimated by \citet{webb03}.
Since the CIRB intensity at $200\,\mu$m is much smaller than the measured 
peak \citep[$\nu I_\nu =15\,\mbox{nW\,m}^{-2} \mbox{sr}^{-1}$ at $\lambda 
\simeq 140\,\mu$m)][]{lagache99}, it strongly supports the existence of 
heavily obscured dusty galaxy population. 
Therefore, LBGs are not the dominant population contributing to the CIRB.

Straightforward conversion of the UV radiation from galaxies into the IR
by recent models cannot reproduce the strong peak \citep[e.g.,][]{balland03}.
In addition, the large energy budget radiated in the IR requires an effective
energy release, which may related to starburst phenomena, even if they
are nearby galaxies \citep{takeuchi01a,franceschini01,hauser01}.
Therefore, hidden starbursts are necessary to explain the strong
evolution of submm source counts \cite[e.g.,][]{takeuchi01a,takeuchi01b}.
At present, redshifts of the starburst population contributing
to the CIRB are almost unknown.
If the dusty starburst is a low-$z$ phenomenon, there must be a strongly
obscured era at $z \sim 1$ \citep{takeuchi01a}.
In this case it is interesting to discuss a possible relation between
the activation of star formation and the peak of merger rate in hierarchical
structure formation scenarios \citep[e.g.,][]{lacey93,kitayama96}.
On the other hand, if the obscured galaxies reside in a high-$z$ Universe,
it implies that the high-$z$ cosmic SFR density may be higher than the 
dust-corrected value \citep{steidel99} because it would be a sum of 
the contributions both from LBGs and hidden starbursts.

\section{Summary and conclusion}\label{sec:conclusion}

Lyman-break galaxies (LBGs) contain a non-negligible amount of dust.
\citet{takeuchi03a} (T03) constructed a model of the infrared spectral 
energy distribution (SED) for very young galaxies by taking into 
account the dust size distribution in the early stage of galaxy evolution,
which can be different from that of present-day evolved galaxies.

In this work, we applied T03 model to LBGs and constructed their expected SED.
In order to examine the grain size distribution of dust, we calculated
the SEDs based on two distinct type of the distribution models: 
a single-sized distribution predicted by \citet{todini01}, and a power-law 
distribution with a slope of $dN/da \propto a^{-3.5}$, which is 
often used to describe the Galactic dust \citep{mathis77}.

We found that the single-sized and power-law dust size distributions yield a
very similar detectability of LBGs at the submillimetre (submm).
We also found that the difference of the grain size distribution affects
the SED drastically at mid-infrared (MIR) wavelengths.
Galaxies with power-law dust distribution have much less flux at MIR
than the other.
Generally, strong outflow is observed in LBGs, and consequently their 
geometry of dust configuration might be relatively simple.
It reduces the uncertainty of complex radiative transfer, and we can 
safely explore the dust grain size distribution high-redshift galaxies 
by (observer-frame) FIR observations.

Then, we applied the model to a gravitationally lensed LBG \cb\ (cB58),
a unique probe of the dust emission from LBGs.
Thanks to the large magnification factor, the dust emission has been 
detected in the submm by SCUBA observations.
These observations suggest that the dust is hot in this galaxy.
Our model framework well reproduced the hot dust temperature under a natural
assumption for this galaxy.

We also examined the detectability of LBGs at submm wavelengths
in an eight-hour deep survey by ALMA.
The LBG population with an age $\ga 10^8\,{\rm yr}$ and
a SFR $\ga 10\,M_\odot {\rm yr}^{-1}$ can be detected in such a survey.
The detectability of LBGs in the submm is not strongly affected by redshift,
because of the well-known negative $K$-correction.
Hence, the detected LBG sample will be undoubtedly an ideal sample to study
the early evolution of metal and dust content in galaxies.

By integrating over their redshifted SEDs with the observed luminosity 
functions, we obtained the contribution of LBGs to the cosmic infrared 
background radiation (CIRB).
Although they have non-negligible amount of dust, their contribution
was found to be small, especially in the FIR $\sim 200\,\mu$m.
Thus, we need a strongly obscured population of galaxies which contains
a large amount of star formation, at some epoch in the history of the universe.

\begin{acknowledgements}
First we deeply thank the anonymous referee whose careful comments 
improved the quality and clarity of this paper very much.
This work has been motivated by a stimulating discussion with Marcin Sawicki
at a conference in Mykonos island.
Akio K.\ Inoue is kindly thanked for his careful reading of the manuscript.
We are greatly indebted to Hiroyuki Hirashita, Veronique Buat, Masato Onodera,
Leslie K.\ Hunt, Andrea Ferrara, Atsunori Yonehara, Tracy M.\ Webb, 
Kouichiro Nakanishi, and Ichi Tanaka
for their helpful comments and suggestions.
\end{acknowledgements}


\begin{thebibliography}{}
\bibitem[\protect\citeauthoryear{Adelberger \& Steidel}{2000}]{adelberger00}
 Adelberger, K.\ L., \& Steidel, C.\ C.\ 2000, ApJ, 544, 218

\bibitem[\protect\citeauthoryear{Balland, Devriendt, \& Silk}
{Balland et al.}{2003}]{balland03}
 Balland, C., Devriendt, J.\ E.\ G., \& Silk, J.\ 2003, MNRAS, 343, 107

\bibitem[\protect\citeauthoryear{Baker et al.}{2001}]{baker01}
 Baker, A.\ J., Lutz, D., Genzel, R., Tacconi, L. J., \& Lehnert, M.\ D.\ 
 A\&A, 372, L37

\bibitem[\protect\citeauthoryear{Baker et al.}{2004}]{baker04}
 Baker, A.\ J., Tacconi, L.\ J., Genzel, R., Lehnert, M.\ D., \& Lutz, D.\
 2004, ApJ, 604, 125

\bibitem[\protect\citeauthoryear{Bechtold et al.}{1997}]{bechtold97}
 Bechtold, J., Yee, H.\ K.\ C., Elston, R, \& Ellingson, E.\ 1997,
 ApJ, 477, L29

\bibitem[\protect\citeauthoryear{Buat et al.}{2002}]{buat02}
 Buat, V., Boselli, A., Gavazzi, G., \& Bonfanti, C.\ 2002, A\&A, 383, 801

\bibitem[\protect\citeauthoryear{Calzetti, Kinney, \& Storchi-Bergmann}
{Calzetti et al.}{1994}]{calzetti94}
 Calzetti, D., Kinney, A., \& Storchi-Bergmann, T.\ 1994, ApJ, 429, 582

\bibitem[\protect\citeauthoryear{Calzetti et al.}{2000}]{calzetti00}
 Calzetti, D., Armus, L., Bohlin, R.\ C., et al.\ 2000, ApJ, 533, 682

\bibitem[\protect\citeauthoryear{Calzetti}{2001}]{calzetti01}
 Calzetti, D.\ 2001, PASP, 113, 1449

\bibitem[\protect\citeauthoryear{Cardelli, Clayton, \& Mathis}
{Cardelli et al.}{1989}]{cardelli89}
 Cardelli, J.\ A., Clayton, G.\ C., \& Mathis, J.\ S.\ 1989, ApJ, 345, 245

\bibitem[\protect\citeauthoryear{Chapman et al.}{2000}]{chapman00}
 Chapman, S.\ C., Scott, D., Steidel, C.\ C.\ et al.\ 2000, MNRAS, 319, 318

\bibitem[\protect\citeauthoryear{Chapman et al.}{2002}]{chapman02}
 Chapman, S.\ C., Shapley, A., Steidel, C., \& Windhorst, R.\ 2002,
 ApJ, 572, L1

\bibitem[\protect\citeauthoryear{Dickinson}{2000}]{dickinson00}
 Dickinson, M.\ 2000, Phil.\ Trans.\ R.\ Soc.\ Lond.\ A, 358, 2001

\bibitem[\protect\citeauthoryear{Dole, Lagache, \& Puget}
{Dole et al.}{2003}]{dole03}
 Dole, H., Lagache, G., \& Puget, J.-L.\ 2003, ApJ, 585, 617

\bibitem[\protect\citeauthoryear{Draine \& Anderson}{1985}]{draine85}
 Draine, B.\ T., \& Anderson, L.\ 1985, ApJ, 292, 494

\bibitem[\protect\citeauthoryear{Draine \& Lee}{1984}]{draine84}
 Draine, B.\ T., \& Lee, H.\ M.\ 1984, ApJ, 285, 89

\bibitem[\protect\citeauthoryear{Draine \& Li}{2001}]{draine01}
 Draine, B.\ T., \& Li, A.\ 2001, ApJ, 551, 807

\bibitem[\protect\citeauthoryear{Dunne \& Eales}{2001}]{dunne01}
 Dunne, L., \& Eales, S.\ A.\ 2001, MNRAS, 327, 697

\bibitem[\protect\citeauthoryear{Dwek \& Scalo}{1980}]{dwek80}
 Dwek, E., \& Scalo, J.\ M.\ 1980, ApJ, 239, 193

\bibitem[\protect\citeauthoryear{Edge et al.}{1999}]{edge99}
 Edge, A.\ C., Ivison, R.\ J., Smail, I., Blain, A.\ W., \& Kneib, J.-P.\ 
 1999, MNRAS, 306, 599

\bibitem[\protect\citeauthoryear{Ellingson et al.}{1996}]{ellingson96}
 Ellingson, E., Yee, H.\ K.\ C., Bechtold, J., \& Elston, R.\ 1996,
 ApJ, 466, L71

\bibitem[\protect\citeauthoryear{Erb et al.}{2003}]{erb03}
 Erb, D.\ K., Shapley, A.\ E., Steidel, C.\ C., et al.\ 2003, ApJ, 591, 101

\bibitem[\protect\citeauthoryear{Fixsen et al.}{1998}]{fixsen98}
 Fixsen, D.\ J., Dwek, E., Mather, J.\ C., Bennett, C.\ L., \& 
 Shafer, R.\ A.\ 1998, ApJ, 508, 123

\bibitem[\protect\citeauthoryear{Franceschini et al.}{2001}]{franceschini01}
 Franceschini, A., Aussel, H., Cesarsky, C. J., Elbaz, D., \& 
 Fadda, D.\ 2001, A\&A, 378, 1

\bibitem[\protect\citeauthoryear{Galliano et al.}{2003}]{galliano03}
 Galliano, F., Madden, S.\ C., Jones, A.\ P., et al.\  2003, A\&A, 407, 159

\bibitem[\protect\citeauthoryear{Gispert, Lagache, \& Puget}
{Gispert et al.}{2000}]{gispert00}
 Gispert, R., Lagache, G., \& Puget, J.\ L.\ 2000, A\&A, 360, 1

\bibitem[\protect\citeauthoryear{Gordon et al.}{2000}]{gordon00}
 Gordon, K.\ D., Clayton, G.\ C., Witt, A.\ N., \& Misselt, K.\ A.\ 2000, 
 ApJ, 533, 236

\bibitem[\protect\citeauthoryear{Hauser \& Dwek}{2001}]{hauser01}
 Hauser, M.\ G., \& Dwek, E.\ 2001, ARA\&A, 39, 249

\bibitem[\protect\citeauthoryear{Hauser et al.}{1998}]{hauser98}
 Hauser, M.\ G., Arendt, R.\ G., Kelsall, T., et al.\ 1998, ApJ, 508, 25

\bibitem[\protect\citeauthoryear{Hirashita, Hunt, \& Ferrara}
{Hirashita et al.}{2002}]{hirashita02} 
 Hirashita, H., Hunt, L.\ K., \& Ferrara, A.\ 2002, MNRAS, 330, L19 (H02)

\bibitem[\protect\citeauthoryear{Hirashita, Buat, \& Inoue}
{Hirashita et al.}{2003}]{hirashita03} 
 Hirashita, H., Buat, V., \& Inoue, A.\ K.\ 2003, A\&A, 410, 83

\bibitem[\protect\citeauthoryear{Hunt, Vanzi, \& Thuan}
{Hunt et al.}{2001}]{hunt01}
 Hunt, L.\ K., Vanzi, L., \& Thuan, T.\ X.\ 2001, ApJ, 377, 66

\bibitem[\protect\citeauthoryear{Inoue}{2002}]{inoue02}
 Inoue, A.\ K.\ 2002, ApJ, 570, 688

\bibitem[\protect\citeauthoryear{Ishii, Takeuchi, & Sohn}
{Ishii et al.}{2002}]{ishii02}
 Ishii, T.\ T., Takeuchi, T.\ T., \& Sohn, J.-J.\ 2002, in Infrared and
 Submillimeter Space Astronomy, EDP Sciences, Les Ulis, p.169

\bibitem[\protect\citeauthoryear{James et al.}{2002}]{james02}
 James, A., Dunne, L., Eales, S., \& Edmunds, M.\ G.\ 2002, MNRAS, 335, 753

\bibitem[\protect\citeauthoryear{Kitayama \& Suto}{1996}]{kitayama96}
 Kitayama, T., \& Suto, Y.\ 1996, MNRAS, 280, 638

\bibitem[\protect\citeauthoryear{Kozasa \& Hasegawa}{1987}]{kozasa87}
 Kozasa, T., \& Hasegawa, H.\ 1987, Prog.\ Theor.\ Phys., 77, 1402

\bibitem[\protect\citeauthoryear{Kr\"ugel}{2003}]{krugel03}
 Kr\"ugel, E.\ 2003, The Physics of Interstellar Dust, Institute of Physics
 Publishing, Bristol

\bibitem[\protect\citeauthoryear{Lacey \& Cole}{1993}]{lacey93}
 Lacey, C., \& Cole, S.\ 1993, MNRAS, 262, 627

\bibitem[\protect\citeauthoryear{Lagache et al.}{1999}]{lagache99}
 Lagache, G., Abergel, A., Boulanger, F., D\'esert, F.\ X., \& Puget, J.-L.\
 1999, A\&A, 344, 322

\bibitem[\protect\citeauthoryear{Lagache, Dole, \& Puget}
{Lagache et al.}{2003}]{lagache03}
 Lagache, G., Dole, H., \& Puget, J.-L.\ 2003, MNRAS, 338, 555

\bibitem[\protect\citeauthoryear{Madden}{2000}]{madden00} 
 Madden, S.\ C.\ 2000, NewAR, 44, 249

\bibitem[\protect\citeauthoryear{Mathis, Rumpl, \& Nordsieck}
{Mathis et al}{1977}]{mathis77}
 Mathis, J.\ S., Rumpl, W., \& Nordsieck, K.\ H.\ 1977, ApJ, 217, 425

\bibitem[\protect\citeauthoryear{Matteucci \& Pipino}{2002}]{matteucci02} 
 Matteucci, F., \& Pipino, A.\ 2002, ApJ, 569, L69

\bibitem[\protect\citeauthoryear{Nakanishi et al.}{1997}]{nakanishi97}
 Nakanishi, K., Ohta, K., Takeuchi, T.\ T., et al.\ 1997, PASJ, 49, 535

\bibitem[\protect\citeauthoryear{Nozawa et al.}{2003}]{nozawa03}
 Nozawa, T., Kozasa, T., Umeda, H., Maeda, K., \& Nomoto, K.\ 2003,
 ApJ, 598, 785

\bibitem[\protect\citeauthoryear{Ouchi et al.}{1999}]{ouchi99}
 Ouchi, M., Yamada, T., Kawai, H., \& Ohta, K.\ 1999, ApJ, 517, L19

\bibitem[\protect\citeauthoryear{Pettini et al.}{2000}]{pettini00}
 Pettini, M., Steidel, C.\ C., Adelberger, K.\ L., Dickinson, M., \& 
 Giavalisco, M.\ 2000, ApJ, 528, 96

\bibitem[\protect\citeauthoryear{Pettini et al.}{2003}]{pettini03}
 Pettini, M., Rix, S., Steidel, C.\ C., et al.\ E.\ 2003, ApJ, 569, 742

\bibitem[\protect\citeauthoryear{Plante \& Sauvage}{2002}]{plante02}
 Plante, S., \& Sauvage, M.\ 2002, AJ, 124, 1995

\bibitem[\protect\citeauthoryear{Sawicki}{2001}]{sawicki01}
 Sawicki, M.\ 2001, AJ, 121, 2405 (S01)

\bibitem[\protect\citeauthoryear{Sawicki \& Yee}{1998}]{sawicki98}
 Sawicki, M., \& Yee, H.\ K.\ C.\ 1998, AJ, 115, 1329

\bibitem[\protect\citeauthoryear{Seitz et al.}{1998}]{seitz98}
 Seitz, S., Saglia, R.\ P., Bender, R., et al.\ 1998, MNRAS, 298, 945

\bibitem[\protect\citeauthoryear{Shapley et al.}{2001}]{shapley01}
 Shapley, A.\ E., Steidel, C.\ C., Adelberger, K.\ L., et al.\ 2001, 
 ApJ, 562, 95

\bibitem[\protect\citeauthoryear{Steidel et al.}{1999}]{steidel99}
 Steidel, C.\ C., Adelberger, K.\ L., Giavalisco, M., Dickinson, M., 
 Pettini, M.\ 1999, ApJ, 519, 1

\bibitem[\protect\citeauthoryear{Steidel et al.}{2003}]{steidel03} 
 Steidel, C.\ C., Adelberger, K.\ L., Shapley, A.\ E., et al.\ 2003, 
 ApJ, 592, 728

\bibitem[\protect\citeauthoryear{Takeuchi, Yoshikawa, \& Ishii}
{Takeuchi et al.}{2000}]{takeuchi00}
 Takeuchi, T.\ T., Yoshikawa, K., Ishii, T.\ T.\ 2000, ApJS, 129, 1

\bibitem[\protect\citeauthoryear{Takeuchi et al.}{2001a}]{takeuchi01a}
 Takeuchi, T.\ T., Ishii, T.\ T., Hirashita, H., et al.\ 2001a, PASJ, 53, 37

\bibitem[\protect\citeauthoryear{Takeuchi et al.}{2001b}]{takeuchi01b}
 Takeuchi, T.\ T., Kawabe, R., Kohno, K., et al.\ 2001b, PASP, 113, 586

\bibitem[\protect\citeauthoryear{Takeuchi et al.}{2003a}]{takeuchi03a}
 Takeuchi, T.\ T.,  Hirashita, H., Ishii, T.\ T., Hunt, L.\ K.,  
 Ferrara, A.\ 2003a, MNRAS, 343, 839 (T03)

\bibitem[\protect\citeauthoryear{Takeuchi, Yoshikawa, \& Ishii}
{Takeuchi et al.}{2003b}]{takeuchi03b}
 Takeuchi, T.\ T.,  Yoshikawa, K., Ishii, T.\ T.\ 2003b, ApJ, 587, L89
(erratum: Takeuchi, T.\ T.,  Yoshikawa, K., Ishii, T.\ T.\ 2004, 
ApJ, 606, L171)

\bibitem[\protect\citeauthoryear{Takeuchi \& Ishii}{2004}]{takeuchi04}
 Takeuchi, T.\ T., \& Ishii, T.\ T.\ 2004, ApJ, 604, 40

\bibitem[\protect\citeauthoryear{Teplitz et al.}{2000}]{teplitz00}
 Teplitz, H.\ I., McLean, I.\ S., Becklin, E.\ E., et al.\ 2000, ApJ, 533, L65

\bibitem[\protect\citeauthoryear{Todini \& Ferrara}{2001}]{todini01}
 Todini, P., \& Ferrara, A.\ 2001, MNRAS, 325, 726 (TF01)

\bibitem[\protect\citeauthoryear{Totani \& Takeuchi}{2002}]{totani02}
 Totani, T., \& Takeuchi, T.\ T.\ 2002, ApJ, 570, 470

\bibitem[\protect\citeauthoryear{van der Werf et al.}{2001}]{vanderwerf01}
 van der Werf, P.\ P., Knudsen, K.\ K., Labb\'e, I., \& Franx, M.\ 2001,
 in Deep Millimeter Surveys: Implications for Galaxy Formation and Evolution, 
 ed.\ J.\ D.\ Lowenthal \& D.\ H.\ Hughes
 (Singapore: World Scientific Publishing), 103

\bibitem[\protect\citeauthoryear{Vijh, Witt, \& Gordon}
{Vijh et al.}{2003}]{vijh03}
 Vijh, U.\ P., Witt, A.\ N., \& Gordon, K.\ D.\ 2003, ApJ, 587, 533

\bibitem[\protect\citeauthoryear{Webb et al.}{2003}]{webb03}
 Webb, T.\ M., Eales, S., Foucaud, S., et al.\ 2003, ApJ, 582, 6

\bibitem[\protect\citeauthoryear{Witt et al.}{2000}]{witt00}
 Witt, A.\ N., \& Gordon, K.\ D.\ 2000, ApJ, 528, 799

\bibitem[\protect\citeauthoryear{Whittet}{1992}]{whittet92}
 Whittet, D.\ C.\ B.\ 1992, Dust in the Galactic Environment, IOP, New York

\bibitem[\protect\citeauthoryear{Yee et al.}{1996}]{yee96}
 Yee, H. K. C., Ellingson, E., Bechtold, J., Carlberg, R.\ G., \& 
 Cuillandre, J.-C.\ 1996, AJ, 111, 1783
\end{thebibliography}
\end{document}